\documentclass[11pt,a4paper]{article}
\usepackage[noadjust]{cite}

\usepackage[margin=2.8cm,bottom=3.5cm]{geometry}

\usepackage{amsmath, amssymb}
\usepackage{color}
\pdfoutput=1 

\usepackage{graphicx} 
\usepackage{subcaption}
\usepackage[T1]{fontenc}

\title{Collisions between kinks with long-range tails: a simple and efficient method}

\author{
        Jo\~ao G. F. Campos \\
        Departamento de F\'isica, Universidade Federal de Pernambuco,\\
        Av. Prof. Moraes Rego, 1235, Recife - PE - 50670-901, Brazil\\
        joao.gfcampos@ufpe.br
            \and
        Azadeh Mohammadi\\
        Departamento de F\'isica, Universidade Federal de Pernambuco,\\
        Av. Prof. Moraes Rego, 1235, Recife - PE - 50670-901, Brazil\\
        azadeh.mohammadi@ufpe.br
}

\begin{document} 
\maketitle

\begin{abstract}
We construct initial configurations for the scattering between kinks with long-range tails. For this purpose, we exploit kink solutions in the presence of Bogomol'nyi-Prasad-Sommerfield (BPS)-preserving impurities. This approach offers a highly efficient method and effortless implementation with a negligible computational cost. Our algorithm has a much smaller complexity than the usual minimization method, becoming more than a hundred times faster in some scenarios. Consequently, conducting kink-antikink simulations becomes remarkably straightforward. 
\end{abstract}

\section{Introduction} 

In recent years, significant attention has been devoted to exploring the interactions of kinks, uncovering intriguing phenomena in the process. Notable contributions include the investigation of kink interactions with long-range tails \cite{christov2019kink}, the discovery of spectral walls within kink interactions \cite{adam2019spectral}, and the computation of a collective coordinate model for $\phi^4$ kink interactions \cite{manton2021collective}. These studies have provided valuable insights into non-perturbative aspects of field theories, spontaneous symmetry breaking, and the dynamics of models with effectively one spatial dimension. In particular, kinks with long-range tails emerge in field theories when a certain potential minimum lacks a mass term, the second derivative of the potential at the minimum. Therefore, they are of higher order, making them long-range and, consequently, highly interactive.

Understanding such systems is a challenging research field currently under development (see Ref.~\cite{khare2022kink} and references therein). 
Some pioneering works on kinks with long-range tails include Refs.~\cite{lohe1979soliton, gonzalez1989kinks}. More recently, several families of polynomial and non-polynomial potentials engendering kinks with long-range tails have been listed \cite{khare2014successive, khare2019family, khare2020wide}. Other examples with explicit analytical solutions can be found in Refs. \cite{gomes2012highly, bazeia2018analytical, gani2020explicit, blinov2022kinks}. Interestingly, it was shown in Ref.~\cite{bazeia2023geometrically} that a long-range tail can be induced on a kink by modifying the kinetic term of the corresponding scalar field.

The study of interactions between long-range kinks presents a challenge due to the limitations of usual initial condition approximations. For instance, the additive ansatz suggests that a kink-antikink configuration can be approximated by summing the individual kink profiles when they are sufficiently separated. However, when the kink's tail decays following a power law, neighboring kinks do not exhibit a negligible superposition because a power-law decay lacks a finite range. As a result, the kinks are never truly well-separated. Therefore, the choice of initial conditions in direct numerical simulations of long-range kinks' interactions plays a crucial role in the observed behavior. Initializing the collision with the standard additive ansatz may introduce unwanted initial energy, which converts into radiation. It results in a wrong magnitude and even sign of the force. Hence, it has been demonstrated in previous works Refs.~\cite{manton2019forces, christov2019kink}  that conventional methods for computing the force between long-range kinks must be adapted in such cases.

Performing the numerical simulation of kinks' scattering requires specialized methods in the long-range regime \cite{christov2019long, christov2021kink, campos2021interaction}.
The first correct simulation of the scattering between kinks with long-range tails was performed in Ref.~\cite{christov2019long}. To that end, the authors developed a new method to construct kink-antikink configurations. The correct initial configuration $\phi(x,t=0)$ was found by requiring that it satisfies the static equation of motion as closely as possible. More explicitly, the following functional was minimized at $t=0$ 
\begin{equation}
\label{eq_minI}
    I[\phi]=\lVert \phi^{\prime\prime}-V^\prime(\phi)\rVert_2^2+\text{constraints},
\end{equation}
where $\lVert.\rVert_2^2$ denotes the two-norm squared of the function. The constraints are needed to ensure that the kinks' centers are approximately fixed in the minimization process. The method was shown to work very well for the kinks initially at rest.

In Ref.~\cite{christov2021kink}, the same authors generalized the method for the case where the kinks have finite initial velocity. Now, $\phi$ is obtained by minimizing the following functional
\begin{equation}
\label{eq_minII}
    I_v[\phi]=\lVert (1-v^2)\phi^{\prime\prime}-V^\prime(\phi)\rVert_2^2+\text{constraints},
\end{equation}
where the extra factor accounts for the Lorentz contraction. Then, a second minimization layer for the field $\chi(x,t)=\dot{\phi}(x,t)$ is performed. Namely, they minimized the following functional at $t=0$
\begin{equation}
\label{first_min}
    J_1[\chi]=\lVert \dot{\chi}-\phi^{\prime\prime}+V^\prime(\phi)\rVert_2^2+\text{constraints},
\end{equation}
where $\phi$ is the solution of the first minimization step, given by eq.~\eqref{eq_minII}. As the functional depends on the time derivative, it was necessary to integrate the equations of motion in a small time interval at every minimization step and to take the two-norm. Therefore, it was more computationally costly than the first minimization layer.

Inspired by the works mentioned earlier, we showed in Ref.~\cite{campos2021interaction} that the second layer of minimization could be performed instead by requiring the field to obey the zero mode equation as closely as possible. More formally, the following functional is minimized at $t=0$
\begin{equation}
\label{second_min}
    J_2[\chi]=\lVert(1-v^2)\chi^{\prime\prime}-V^{\prime\prime}(\phi)\chi\rVert_2^2+\text{constraints},
\end{equation}
where $\phi$ is again the solution of the first minimization, given by eq.~\eqref{eq_minII}. It is a generalization of the single kink case, where $\chi$ is proportional to the zero mode, related to the model's translation symmetry. In other words, the field $\chi$ should be proportional to a generalized translation mode for multiple kinks. This method is much less computationally costly than the one presented in Ref.~\cite{christov2021kink} as it does not need integration at every minimization step.

The issue with minimization methods, however, is that they are quite costly in general. Although it is possible to perform them on a large scale \cite{bazeia2023kink}, they require considerable computational time. Here, we propose a new method to construct kink-antikink initial configurations for long-range kinks with negligible computational time. To aim for this purpose, we will exploit impurities that preserve half the system's Bogomol'nyi-Prasad-Sommerfield (BPS) property \cite{adam2019phi4, adam2019spectral}. 

The remaining sections are structured as follows. In section 2, we review the models that preserve half the system's BPS property, and building upon them, we present our new method of initialization for long-range kink-antikink interactions. In section 3, we choose two specific impurities with one and three free parameters to study the collisions in $\phi^8$ and $\phi^{12}$ models. Finally, in section 4, we summarize our ﬁndings.

\section{Half-BPS impurity theories} 

Consider the following scalar field theory in (1+1) dimensions
\begin{equation}
    \mathcal{L}=\frac{1}{2}\phi_t^2-\frac{1}{2}\phi_x^2-V(\phi).
\end{equation}
If the potential is non-negative and contains multiple degenerate vacua, the model exhibits kink solutions $\phi_K(x)$, described by the BPS equation
\begin{equation}
\label{eq:BPS}
\frac{d\phi_K}{dx}=\pm W(\phi_K),
\end{equation}
where $W(\phi)=\sqrt{2V(\phi)}$. We fix the center of the kink $\phi_0$ at $-x_0$, i.e., $\phi_K(-x_0)=\phi_0$. To obtain a kink solution, we assume that $\phi_-<\phi_0<\phi_+$, where $\phi_-$ and $\phi_+$ are two neighboring potential minima. 

Now, consider the related field theory
\begin{equation}
    \mathcal{L}=\frac{1}{2}\phi_t^2-\frac{1}{2}[\phi_x-\sigma(x;\boldsymbol{\alpha}) W(\phi)]^2,
\end{equation}
where $\sigma(x;\boldsymbol{\alpha})$ is an impurity containing $p$ free parameters $\boldsymbol{\alpha}=(\alpha_1,\alpha_2,\cdots,\alpha_p)$. After its inclusion, only one BPS equation is preserved. We are interested in the BPS solution $\phi_{K,\sigma}(x;\boldsymbol{\alpha})$, defined by the following equation
\begin{equation}
\label{eq:BPS_imp}
\frac{d\phi_{K,\sigma}}{dx}=\sigma(x;\boldsymbol{\alpha})W(\phi_{K,\sigma}),
\end{equation}
with the condition $\phi_{K,\sigma}(-x_0;\boldsymbol{\alpha})=\phi_0$. To solve eq.~\ref{eq:BPS_imp} in terms of the $\phi_K(x)$ and $\sigma(x;\boldsymbol{\alpha})$ functions, we change coordinates to \cite{adam2019solvable}
\begin{equation}
\xi(x)=-x_0+\int_{-x_0}^x\sigma(x^\prime;\boldsymbol{\alpha})dx^\prime. 
\end{equation}
Hence, the impurity is removed from the BPS equation, giving rise to
\begin{equation}
    \label{eq:BPS_imp2}
\frac{d\phi_{K,\sigma}}{d\xi}=W(\phi_{K,\sigma}),
\end{equation}
with the boundary condition $\phi_{K,\sigma}(\xi=-x_0;\boldsymbol{\alpha})=\phi_0$. The solutions is readily obtained as $\phi_{K,\sigma}=\phi_K(\xi(x,\boldsymbol{\alpha}))$.
Due to the BPS property, the model possesses a generalized translation symmetry \cite{adam2019phi4} with the associated zero mode $\phi_K^\prime(\xi(x;\alpha))$, where the derivative is with respect to the function's argument.

\textit{The New Idea.} In Ref. \cite{manton2019iterated}, the authors proposed that BPS solutions with specific impurities could describe field configurations containing multiple kinks. We aim to use such profiles as actual initial data to simulate kink collisions. We discovered that $\phi_{K,\sigma}(x;\boldsymbol{\alpha})$ is an excellent approximation for the initial condition, even for the long-range kinks, for some choices of $\sigma$ with an optimal $\boldsymbol{\alpha}$.

Our construction offers two significant improvements. Firstly, only a few parameters require adjustment, which can be accomplished within a negligible computational time. Secondly, there is no need to add any constraint in the minimization function if one chooses $\sigma$ appropriately. The BPS equation already fixes the center of the kink located at $-x_0$, and the center of the opposing kink will automatically be fixed for all $\alpha$ by our choice of the impurity $\sigma$.

The remaining initial condition, now for the velocity field $\dot\phi$, which we have already defined as $\chi$, can be similarly obtained at $t=0$. We consider a family of generalized translation modes of the impurity model. Namely, we have
\begin{equation}
    \chi_{K,\sigma}(x;\boldsymbol{\beta})=-v\phi_K^\prime(\xi(x;\boldsymbol{\beta})).
\end{equation}
For some impurities, the BPS solution has a lump character, resembling a kink-antikink configuration. Changing the integration constant of the BPS solution, which moves the system in the moduli space, can either increase or decrease the lump size, moving the kink and the antikink in opposite directions \cite{adam2023moduli}. Hence, the initial velocity field of the original problem can be approximated remarkably well by $\chi_{K,\sigma}(x;\boldsymbol{\beta})$ for some choices of $\sigma$ with an optimal $\boldsymbol{\beta}$.

It is worth mentioning that there is no need to consider the same impurity for $\phi$ and $\chi$. However, we consider the same impurity $\sigma$ for both for simplicity. The value of the optimal parameters $\alpha$ and $\beta$ are not generally equal, which is the case for the models we study here. 

\section{Kink-Antikink Collisions} 

Let us start with a simple impurity profile, denoted by $\sigma(x;\alpha)=-\tanh(\alpha x)$ with one free parameter $\alpha$. We will refer to the corresponding BPS solution as $\phi_{KA}(x;\alpha)$. The reason for the notation is clear; it closely resembles a kink-antikink configuration \cite{manton2019iterated, adam2020kink}. The qualitative explanation for this property is the following. For large and negative $x$, $\sigma(x,\alpha)\simeq1$ and eq.~\eqref{eq:BPS_imp} becomes the BPS equation for the kink. As $x$ increases, the impurity changes signs, resulting in the BPS equation for the antikink. Moreover, it is easy to see that $\xi(x)$ is an even function. Therefore, the solution is mirrored with respect to the origin, leading to the desired behavior.

Our proposed method involves incorporating a gamma factor into the BPS equation \eqref{eq:BPS_imp}, aiming to include the Lorentz contraction in the collision process. Then, the following function is minimized 
\begin{equation}
\label{eq:min-func}
    F(\alpha)=\left\lVert(1-v^2)\phi_{KA}^{\prime\prime}(\alpha)-V^\prime\left(\phi_{KA}(\alpha)\right)\right\rVert_2^2,
\end{equation} 
where the two-norm is taken only in the interval $[-x_0,x_0]$ and we have omitted the $x$ dependence for conciseness. The extra factor $(1-v^2)$ is the same as the one included in eq.~\eqref{eq_minII}. It is essential for finite initial velocities and is consistent with including a gamma factor in the BPS equation. Then, the optimal $\beta$ is found by minimizing the function
\begin{equation}
    G(\beta)=\left\lVert(1-v^2)\chi_{KA}^{\prime\prime}(\beta)-V^{\prime\prime}\left(\phi_{KA}\right)\chi_{KA}(\beta)\right\rVert_2^2,
\end{equation}
where the $x$ dependence has also been suppressed.

A better fit for the initial field configuration could be obtained by taking impurities with more free parameters simply because there are more parameters to be adjusted. The impurities should be kink-like with asymptotic values $\pm1$ for kink-antikink collisions. Therefore, let us consider the following impurity with three free parameters
\begin{equation}
\label{eq:BPS3}
  \sigma(x;\boldsymbol{\alpha})=-\tanh(\alpha_1x+\alpha_2\tanh(\alpha_3x)). 
\end{equation}
It describes a more general family of kink-like profiles and encompasses the $-\tanh(\alpha x)$ case. This way, we find remarkable results for the $\phi^8$ and $\phi^{12}$ models, as described below.

\subsection{The $\phi^8$ model} 

\begin{figure}
    \centering
    \includegraphics[width=0.85\textwidth]{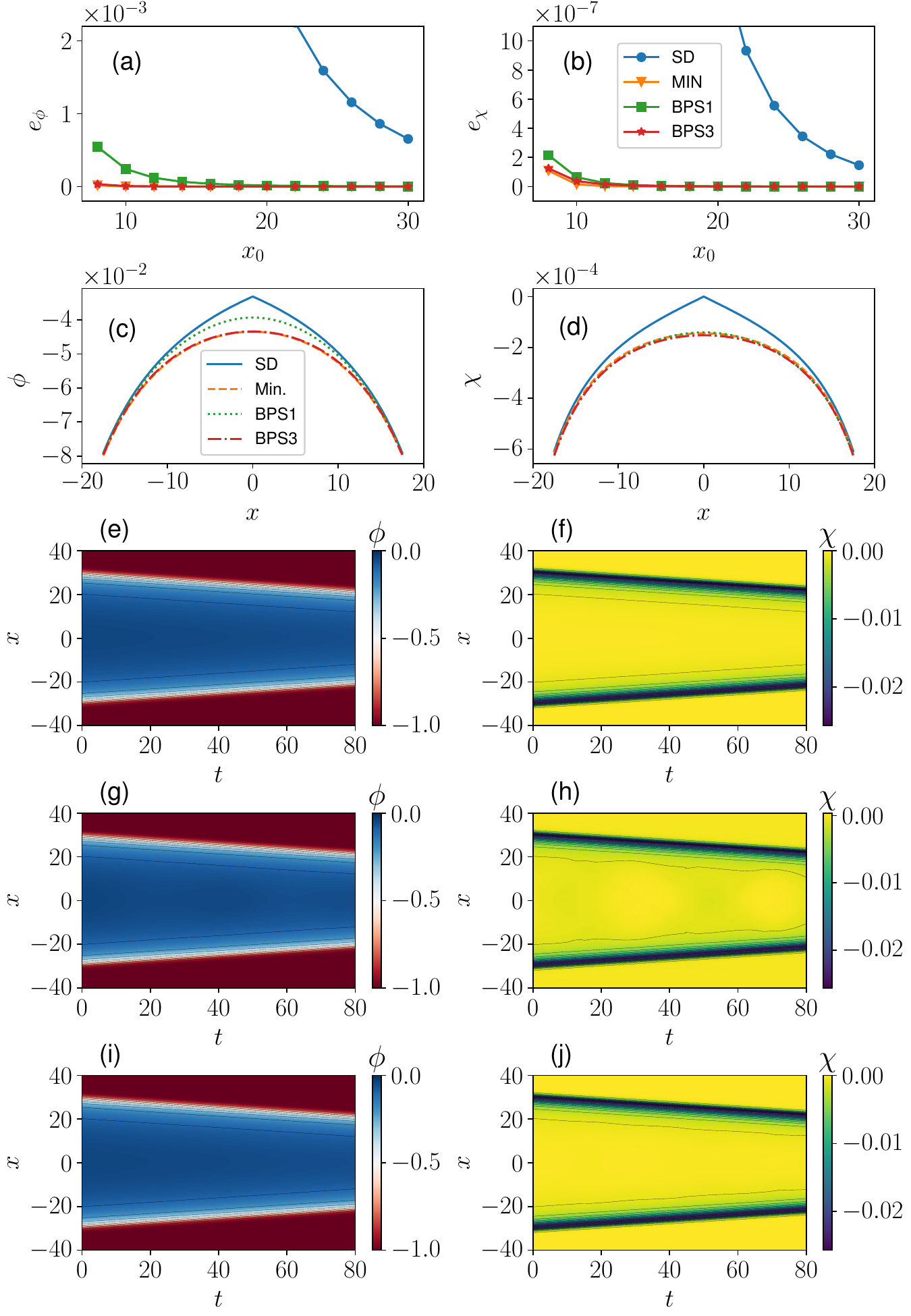}
    \caption{(a-b): Values of the $e_\phi$ and $e_\chi$ as a function of $x_0$ for several methods. (c-d) The initial condition for $\phi$ and $\chi$. Evolution of the fields in spacetime for the minimized method (e-f), the BPS1 method (g-h), and the BPS3 (i-j). We fix $v=0.1$.}
    \label{fig1}
\end{figure}

In order to test the method described above, we considered the following $\phi^8$ potential
\begin{equation}
    V(\phi)=\frac{1}{2}\phi^{2n}(1-\phi^2)^2,
\end{equation}
where $n=2$. The model has two asymmetric kink solutions $\phi_{-1,0}$ and $\phi_{0,+1}$ and the corresponding antikink solutions. The tails $\phi\to \pm 1$ go exponentially in the form $e^{-2|x|}$ to the asymptotic values $\pm1$, and the one $\phi\to 0$ converges in the power-law form $1/x$ to the asymptotic value $0$, hence long-range in this tail. In fact, the tail $\phi \to 0$ is long-range for $n\geq 2$ considering the above potential, with power-law behavior given by $x^{-1/(n-1)}$. We construct kink-antikink configurations where the long-range tails are superposing. We denote the two impurity models $-\tanh(\alpha x)$ and eq.~\eqref{eq:BPS3} as $BPS1$ and $BPS3$, respectively. The Euclidean norm of the differential equation at the optimal parameter, $F(\boldsymbol{\alpha}_o)\equiv e_\phi$ and $G(\boldsymbol{\beta}_o)\equiv e_\chi$, are shown in Figs.~\ref{fig1}(a) and \ref{fig1}(b) as a function of the half-separation $x_0$, fixing $v=0.1$. As a comparison, the norm of the split-domain (SD) ansatz and the usual minimized (MIN) solutions are also shown. The SD solution is the more accurate among the naive initial conditions, such as the additive ansatz and product ansatz, while the MIN solution gives the reference value of the field. 

The BPS1 solution performs much better than the SD one, being much closer to the MIN. As expected, all solutions converge to the same values as $x_0$ increases because the kink-antikink superposition decreases. Remarkably, the BPS3 solution has an excellent agreement with the MIN, with the two curves coinciding in the scale shown in the graph. This can also be observed in the field profiles near the origin in Figs.~\ref{fig1}(c) and \ref{fig1}(d).  

Let us compare the actual evolution of the field in spacetime for $x_0=30$ and $v=0.1$. Figs.~\ref{fig1}(e) and \ref{fig1}(f) show the MIN solution, which evolves smoothly, as expected. In the BPS1 solution, shown in Figs.~\ref{fig1}(g) and \ref{fig1}(h), we find a very similar evolution, but it is possible to see small oscillations in the inner contour. The observed deviation is indeed very small. The contour plot in black shows the field's fine scale; without it, it would be impossible to tell the difference between the graphs. Finally, the BPS3 solution is shown in Figs.~\ref{fig1}(i) and \ref{fig1}(j). It exhibits remarkable similarity with the MIN solution.

It is possible to estimate the magnitude of the contour oscillations in the field evolution comparing the BPS method and the MIN one as a reference. Using the data in Fig.~\ref{fig1}, we obtain errors of order $10^{-3}$ and $10^{-4}$, for $\phi$ and $\chi$ respectively, considering  the BPS1 method. For the BPS3, they are of order $10^{-4}$ and $10^{-5}$, i.e., roughly 10 times more accurate than the BPS1.

\begin{figure}
    \centering
    \includegraphics[width=0.85\textwidth]{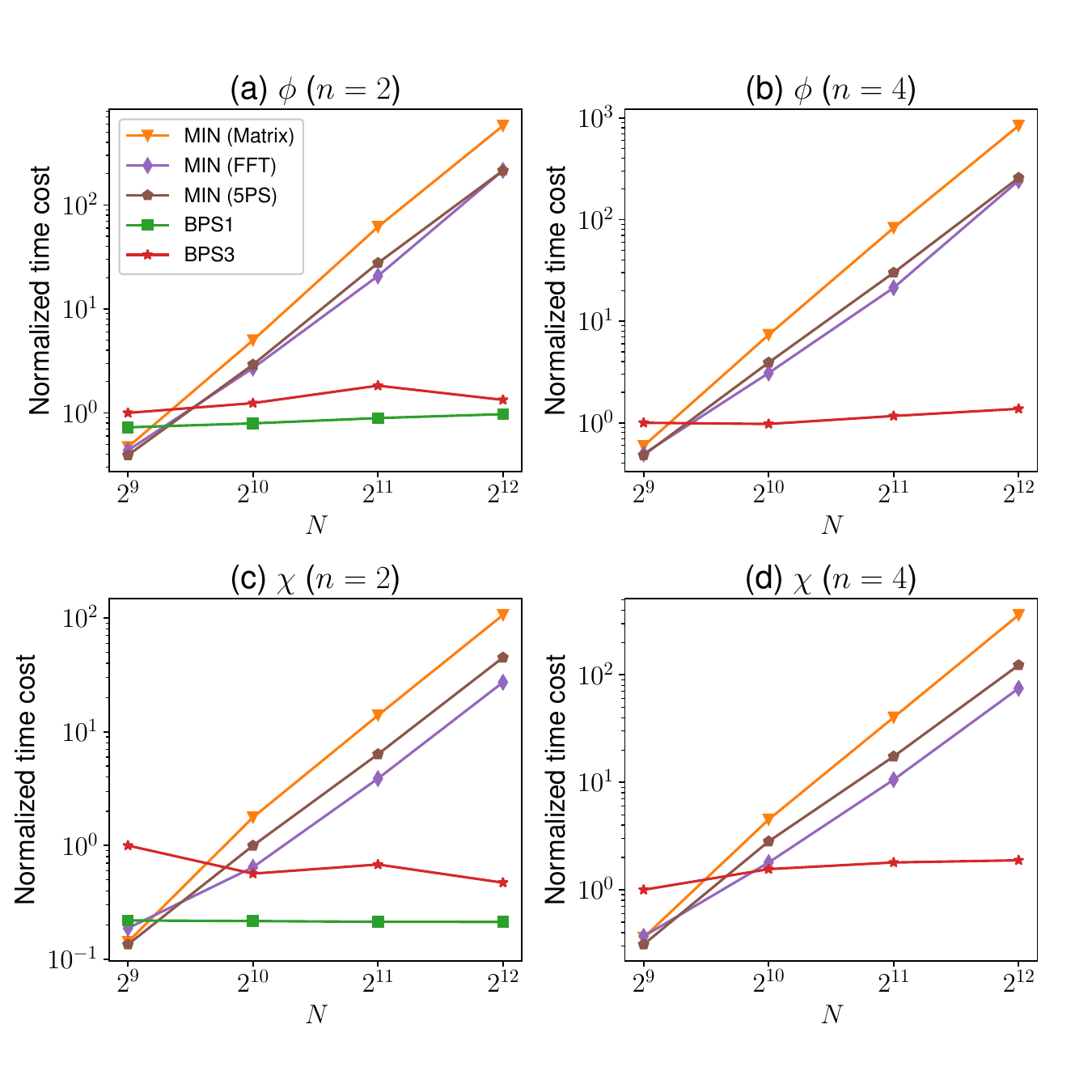}
    \caption{Normalized time cost of the first minimization step for several minimization schemes considering (a) and (c) the $\phi^8$ model, and (b) and (d) the $\phi^{12}$ model. Initial guesses are: (a) $\alpha=1.0$ and $\boldsymbol{\alpha}=(\alpha_1,\alpha_2,\alpha_3)=(8.49,-1109,0.00757)$ for BPS1 and BPS3, respectively; (b) $\boldsymbol{\alpha}=(\alpha_1,\alpha_2,\alpha_3)=(4.14,-1109,0.0037)$; (c) $\beta=0.1$ and $\boldsymbol{\beta}=(\beta_1,\beta_2,\beta_3)=(0.6,-9,0.037)$ for BPS1 and BPS3, respectively; (d) $\boldsymbol{\beta}=(\beta_1,\beta_2,\beta_3)=(0.39,-0.84,0.26)$.}
    \label{fig:complexity}
\end{figure}

We estimate the algorithmic complexity of the BPS method, which is how the execution time scales with the size of the data set, in the following way. To mimic the usual simulation procedure in the literature, we fix $x_0=25.0$ and perform several minimizations with different $v$. We pick eight equally spaced velocities in the interval $[0.1,0.8]$. They are picked in an increasing fashion, using the previous result as an initial guess for the next minimization. The box is fixed at the interval $[-100.0,100.0]$. The time execution of the first minimization step as a function of the number of mesh points N, which is the size of our data set, is shown in Fig.~\ref{fig:complexity}(a) for several methods. 

The MIN method can be performed using several derivative approximations. We utilized the pseudospectral method, which can be performed either via matrix multiplication or the Fast Fourier Transform (FFT) \cite{trefethen2000spectral}. Finally, we considered the five-point stencil (5PS) approximation for the derivative. The 5PS and the pseudospectral using the FFT have better performance. The former scales roughly as $N^3$, while the latter does not fit into a power-law behavior because the algorithmic complexity of the FFT contains a $\log(N)$ factor. On the other hand, the BPS methods scale roughly as $N^0$, offering a significant time gain for sufficiently large $N$. The time cost of the BPS methods does not increase with $N$ because the most costly step in the process is, in fact, solving the BPS equation, not taking the derivative. 

Remarkably, the BPS3 minimization is at least 160 times faster than the usual minimization for $N=4096$. This gain is enough to allow simulations on a moderate scale to be performed even on a personal computer. It is worth mentioning that a good initial guess for $\alpha$ and $\beta$ should be provided to the BPS methods in order to obtain fast and correct results. We start with a simple guess in the first run, such as $\boldsymbol{\alpha}=(1.0,1.0,1.0)$. After a few minimizations, the result converges to an optimal range of values. Then, we have a much better range of initial guesses for subsequent minimizations, leading to a better performance.

The second minimization step has similar scaling behavior; see Fig.~\ref{fig:complexity}(c). However, the relative time gain is not as significant as in the first minimization step because minimizing $\chi$ is the less costly step when it is performed the usual way. The reason is simple. Usually, the field $\chi$ is less long-range than $\phi$ because it is related to the zero mode, that is, the derivative of the kink configuration in space. Thus, we have, for instance, that the kink asymptotic behavior in the $\phi^{8}$ model is proportional to $x^{-1}$, whereas the zero mode asymptotic behavior is proportional to $x^{-2}$. Therefore, minimizing $\chi$ using the usual method is much easier for power-law tails in general, making the advantage of BPS construction more modest. Nonetheless, in absolute terms, the BPS minimization of both $\phi$ and $\chi$ have small and comparable time costs.

\begin{figure}
    \centering
    \includegraphics[width=0.85\textwidth]{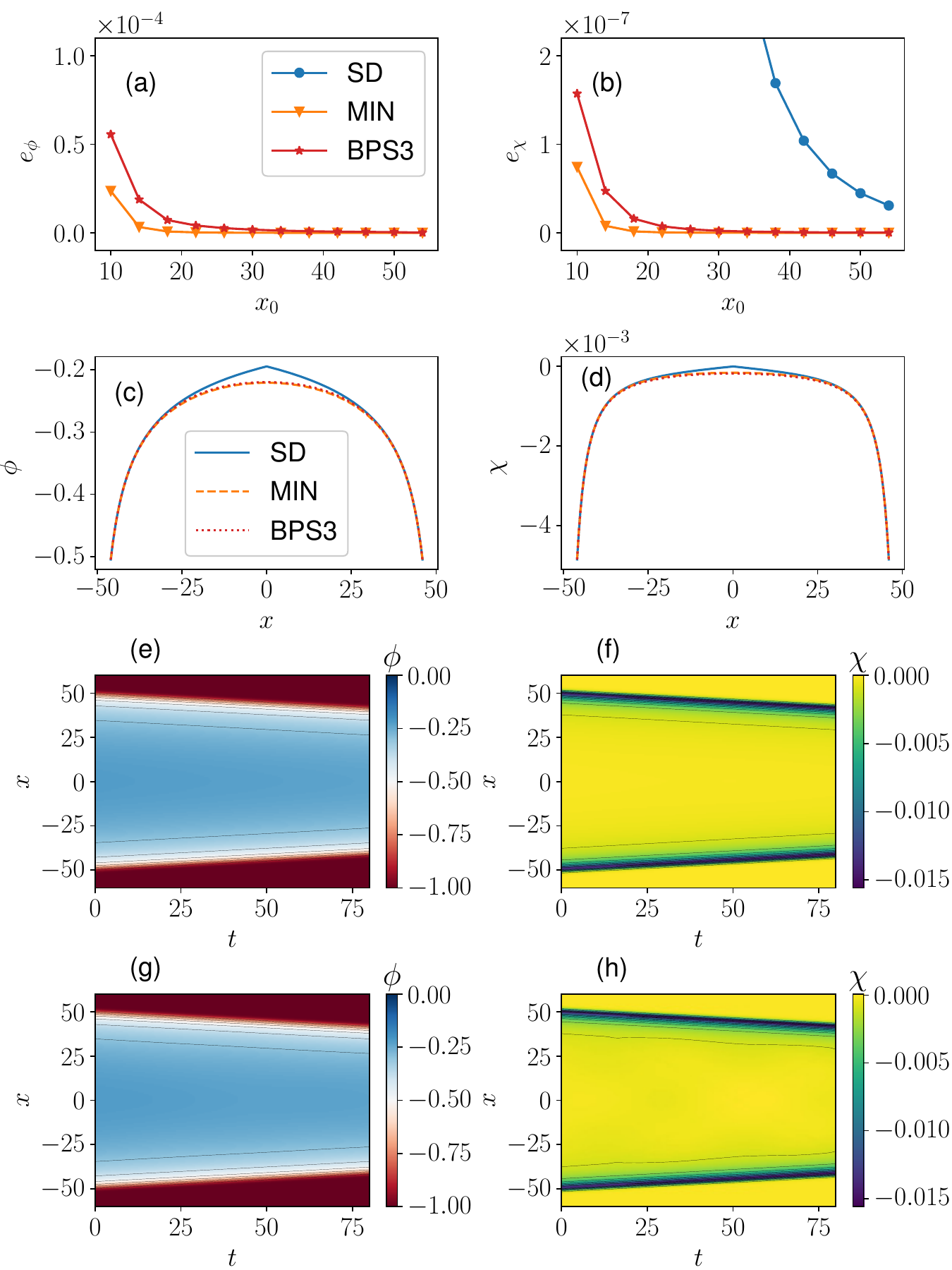}
    \caption{(a-b): Values of the $e_\phi$ and $e_\chi$ as a function of $x_0$ for several methods. The SD error does not appear in the graph scale of (a). (c-d) The initial condition for $\phi$ and $\chi$. Evolution of the fields in spacetime for the minimized method (e-f) and the BPS3 method (g-h). We fix $v=0.1$.}
    \label{fig2}
\end{figure}

We repeated the procedure for several models, including both the $\phi^{10}$ ($n=3$) and $\phi^{12}$ ($n=4$) models. We found that the BPS3 is an excellent method to find minimized initial conditions in general. In order to illustrate this point, we report our results for the $\phi^{12}$ ($n=4$), which contain very fat tails, i.e., with $x^{-1/3}$ asymptotic form. 

\subsection{The $\phi^{12}$ model} 

To assess the accuracy of the BPS3 method for the $\phi^{12}$ $(n=4)$ model, we compare it with the MIN solution in Fig.~\ref{fig2} fixing $v=0.1$. The error functions $e_\phi$ and $e_\chi$ are shown in Figs.~\ref{fig2}(a) and \ref{fig2}(b). Again, the BPS error is very close to the reference value, both converging as $x_0$ increases. Fixing $x_0=50.0$ and $v=0.1$, we obtain an excellent agreement of the initial conditions near the origin (see Figs.~\ref{fig2}(c) and \ref{fig2}(d)), where the superposition occurs. The evolution of the fields in spacetime is shown in Figs.~\ref{fig2}(e) and \ref{fig2}(f) for the MIN initial condition, and in Figs.~\ref{fig2}(g) and \ref{fig2}(h) for the BPS3. They are almost indistinguishable, except for a small deviation in the contour closest to the origin. Now, the relative errors for $\phi$ and $\chi$ are of order $10^{-3}$ and $10^{-4}$, respectively. 

We repeat the procedure to estimate the time cost of the first minimization step with $x_0=50.0$ and obtain roughly the same scaling behavior. The result is shown in Fig.~\ref{fig:complexity}(b) and (d) for $\phi$ and $\chi$, respectively. Again, remarkably, the BPS3 method is at least 175 times faster than the usual minimization For $N=4096$. The previous considerations for the $\chi$ minimization also apply to the $\phi^{12}$ model.

It is possible to obtain even more significant time gains by considering BPS impurities with two free parameters. One interesting approach is to do a few tests to find the average values of $\alpha_i$ in eq.~\eqref{eq:BPS3}, then fix the one with a smaller variance to its average value. This procedure increases the time gain without losing significant accuracy. 

\begin{figure}
    \centering
    \includegraphics[width=0.85\textwidth]{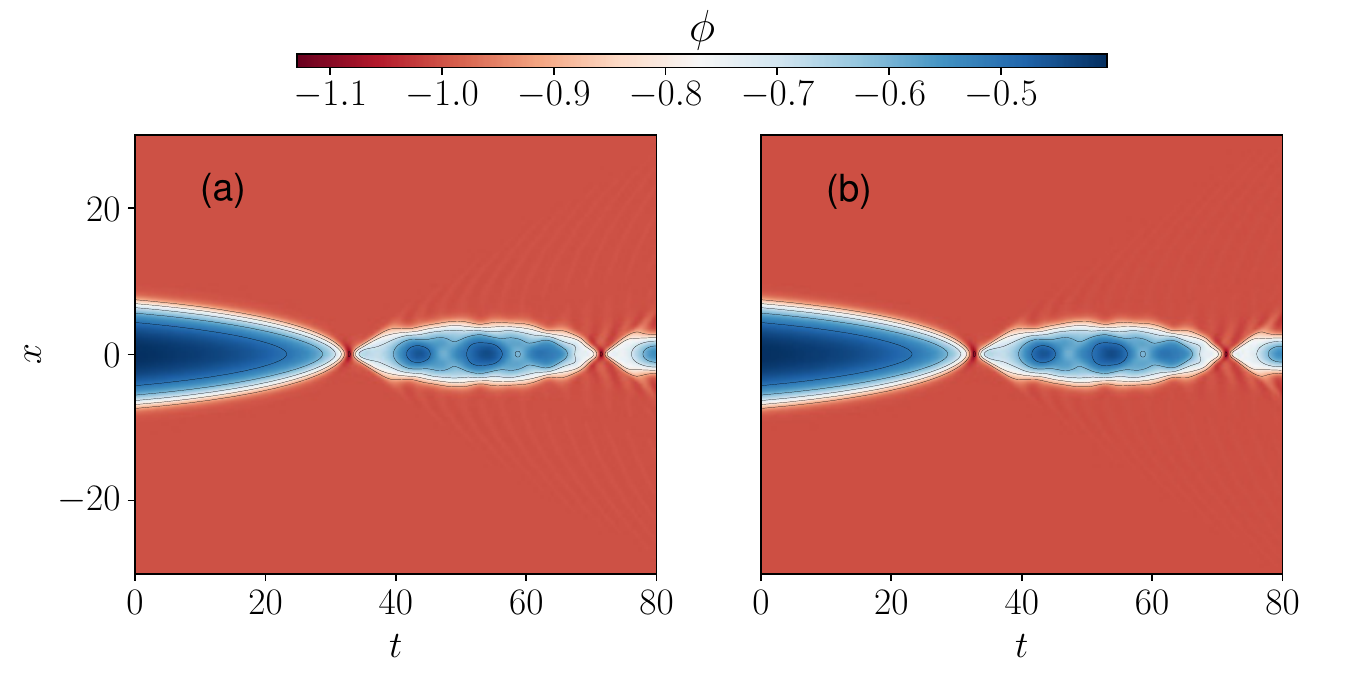}
    \caption{Field evolution in spacetime for the initial configuration taking (a) the usual minimization method and (b) the BPS minimization with the three-parameter impurity BPS3. Parameters are $x_0=10.0$, $v=0.1$, and $n=4$.}
    \label{fig3}
\end{figure}

To put the accuracy of our construction to a final test, we consider a relatively small half distance $x_0=10.0$ and perform a numerical simulation using the BPS3 initial condition and the usual minimization method as a reference. The result is shown in Fig~\ref{fig3} for $v=0.1$. We present the reference method in panel (a) and BPS3 in panel (b). Even though there is a deviation from the reference solution in the previous graphs, it is mainly focused on the evolution before the kinks superpose. The error is very small compared to the field variation near the bounce. Therefore, it is effectively erased, and the collision output is virtually identical to the reference one.

\section{Conclusion} 

In this paper, we offered a construction to find initial conditions that generate smooth field evolution of kink-antikink configurations with long-range tails. It was achieved with the aid of half-BPS preserving impurities. By considering kink-like impurities, we obtained configurations with a kink-antikink character similar to the ones discussed in  Ref.~\cite{manton2019iterated}. The optimal value of the impurity parameter was obtained by virtue of the minimization of appropriate functions. For the component $\phi$, the function was the two-norm of the static equation of motion, and for the component $\chi$, the two-norm of the zero mode equation. Then, an initial condition was obtained with remarkable similarity with the reference method. 

The advantages of our construction are that the constraints are always guaranteed by the chosen impurities, and only a few parameters need to be optimized, a task that can be performed in a negligible time in modern computers. It implies that our method has a negligible increase in the time cost with the number of mesh points $N$. Thus, for the component $\phi$, it becomes more than a hundred times faster than the usual minimization for $N=4096$ in both models considered here. Moreover, by letting the kinks move closer to each other, we see that the error in our approximation is indeed much smaller than the variation of the field at the bounces, having an unnoticeable effect in the time evolution for moderate time intervals.

Our idea may be generalized to find multi-soliton configurations with long-range tails in two or higher dimensions. More importantly, it also makes the simulation of long-range kinks much more accessible, boosting scientific advances in the field. Accordingly, a natural continuation of the present work is applying our method for large-scale simulations of interactions between kinks with long-range tails.

\section*{Acknowledgments}
A. M. acknowledges financial support from the National Council for Scientific and Technological Development - CNPq, Grant no. 309368/2020-0, the Brazilian agency CAPES and also Universidade Federal de Pernambuco Edital Qualis A. J. G. F. C. acknowledges financial support from CNPq, Grant no. 150166/2022-2, and the Brazilian agency FACEPE, Grant no. BFP-0013-1.05/23.

\providecommand{\noopsort}[1]{}\providecommand{\singleletter}[1]{#1}%

\end{document}